\documentclass{article}
\usepackage{amsmath}
\usepackage{graphicx}
\usepackage{amssymb}
\usepackage{multirow}
\usepackage{array}
\usepackage{color}
\usepackage{epsfig} 
\newtheorem{theorem}{Theorem}

\newtheorem{corollary}[theorem]{Corollary}
\newtheorem{newalgorithm}[theorem]{Algorithm}
\newtheorem{definition}[theorem]{Definition}
\newtheorem{proposition}[theorem]{Proposition}
\newtheorem{example}[theorem]{Example}

\newcommand{\eps}{\varepsilon}

\newcommand{\pwr}{\mbox{${\rm Pwr}$}}

\newcommand{\tH}{\tilde{H}}
\newcommand{\tG}{\tilde{G}}
\newcommand{\tdelta}{\tilde{\delta}}

\newcommand{\txi}{\tilde{\xi}}

\newcommand{\lfteqn}{\vspace{-0.08in} \begin{eqnarray} \begin{array}{lllllll}}
\newcommand{\ndeqn}{\vspace{-0.08in} \end{array} \nonumber \end{eqnarray}}
\newcommand{\Lfteqn}{\vspace{-0.08in} \begin{eqnarray} \begin{array}{lllllll}}
\newcommand{\Ndeqn}{\vspace{-0.08in} \end{array}  \end{eqnarray}}

\title{\bf A Uniform Approach to Maximal Permissiveness
 in Modular Control of Discrete-Event Systems}

\author{Jan Komenda, Feng Lin, and Jan H. van Schuppen 
\thanks{
The work of Jan Komenda is supported by RVO 67985840 and GA{\v C}R
grant 19-06175J. The work of Feng Lin is supported by the National
Science Foundation of USA under Grant 1507096 and by the National
Science Foundation of China under Grants 61673297 and 61773287.}
\thanks{Jan Komenda is with Institute of Mathematics, Academy of Sciences of the Czech
Republic, {\v Z}i{\v z}kova 22, 616 62 Brno, Czech Republic, Feng
Lin is with Department of Electrical and Computer Engineering,
Wayne State University, Detroit, MI 48202, USA, and Jan H. van
Schuppen is with Van Schuppen Control Research, Gouden Leeuw 143,
1103 KB Amsterdam, The Netherlands.}}

\begin{document}
\maketitle \thispagestyle{empty} \pagestyle{empty}

\begin{abstract}

In this paper, a uniform  approach to maximal permissiveness in
modular control of discrete-event systems is proposed. It is based
on three important concepts of modular closed-loops: monotonicity,
distributivity, and exchangeability. Monotonicity of various
closed-loops satisfying a given property  considered in this paper
holds whenever the underlying property is preserved under language
unions. Distributivity holds if the inverse projections of local
plants satisfy the given property with respect to each other.
Among new results, sufficient conditions are proposed for distributed computation
of supremal relatively observable sublanguages.
\end{abstract}

\section{Introduction}

Discrete-event systems (DES) modeling real technological systems
are typically represented as synchronous products of a large
number of relatively small local components modeled as finite
automata \cite{RW89}. In order to guarantee safe operation of the
resulting global system, a safety specification is given and it is
required that the controlled system is included in this
specification. As only controllable (and observable in presence of
partial observations) specifications are achieved, the computation
of sublanguages satisfying these conditions are of paramount
importance. The synthesis of observable sublanguages is difficult,
especially in the modular setting, where the number of states can
grow exponentially with the number of local components.

Unfortunately, observability is not preserved under language
unions, unlike controllability. Therefore, the supremal observable
sublanguage does not always exist, and there are only maximal
observable sublanguages, which are not unique in general. A
stronger notion, called normality, coincides with observability in
the case when all controllable events are observable. Supremal
normal sublanguages exist, but they are difficult to compute,
especially in the modular framework. We have studied possibilities
of local (modular) computations of supremal normal sublanguages in
\cite{KvS08} for local specification languages and in \cite{KvS07}
for global specification languages.

Relative observability was introduced and studied in \cite{RO13}
in the framework of partially observed DES
as a condition stronger than observability and weaker
than normality. It was shown to be closed under language
unions, which makes it an interesting notion that can replace
normality in practical applications.

In this paper, a unifying approach is presented for local
computation of maximally permissive supervisors in modular DES.
This approach is computationally attractive, but the optimality
(supremality of the computed sublanguage) is only guaranteed under
some additional conditions.

We emphasize that in case of global specification languages
coordination control has been proposed in \cite{KMvS12}, which can
reduce the supervisory control problem with global specification
to the case of local specification based on the concept of
conditional decomposability. Therefore we use a more general
framework of coordination control rather than modular control in
this  paper.

In this paper a new approach to maximal permissiveness for modular
and coordination control of DES is presented.
It can be applied to computation of various supremal sublanguages
having properties that are preserved under language unions, in
particular to distributed computation of supremal
normal and relatively observable sublanguages.
Moreover, we show that mutual normality is equivalent to global
mutual normality. Finally we present sufficient conditions for
distributed computation of supremal relatively observable
sublanguages based on the concept of global mutual observability
between the local plants.

The paper is organized as follows. The next section recalls the
basic results of supervisory control theory used further. Section
3 presents three algebraic concepts that enable distributed
computation of supremal sublanguages. In Sections 4, 5, and 6,
sufficient conditions for maximal permissiveness of local control
synthesis are presented.

\section{Preliminary Results} \label{Preliminary}
We first briefly recall some basic notations and concepts in
supervisory control theory of DES. We use an
alphabet $A$ to describe the set of event. The free monoid of
words over $A$ is denoted by $A^*$. Each word is a string of
events. Languages are subsets of $A^*$. The {\em prefix closure}
of a language $L\subseteq A^*$ is $\overline{L}=\{w\in A^* :
(\exists v \in A^*) wv\in L\}$. $L$ is said to be {\em
prefix-closed} if $L=\overline{L}$. We only study prefix-closed
languages in this paper.

A {\em generator\/} is a quadruple $G=(Q,A,\delta,q_0)$,
consisting of a finite set of {\em states} $Q$,  a finite set of
{\em events} $A$, a {\em partial transition function} $\delta: Q
\times A \to Q$, and an {\em initial state}  $q_0 \in Q$. With a
slight abuse of notation, the set of all possible transitions is
also denoted by $\delta$: $\delta = \{ (q, a, q') : \delta (q, a)
= q' \}$. Transition function $\delta$ can be extended in the
standard way to strings, that is, $\delta:Q \times A^*\to Q$. The
{\em language generated} by $G$ is the set of all strings
(trajectories) that can be generated by $G$ and is defined as
$L(G) = \{s\in A^* : \delta(q_0,s)! \}$, where $\delta(q_0,s)!$
means that $\delta(q_0,s)$ is defined.

A {\em controlled generator\/} over an alphabet $A$ is a triple
$(G,A_c,\Gamma)$, where $G$ is a generator over $A$, $A_c\subseteq
A$ is a set of {\em controllable events\/}, $A_{u} = A \setminus
A_c$ is the set of {\em uncontrollable events\/}, and $\Gamma =
\{\gamma \subseteq A : A_{u} \subseteq \gamma\}$ is the {\em set
of control patterns}.

A {\em natural projection} $P: A^* \to B^*$, for $B\subseteq A$,
is a homomorphism defined as $P(a)=\eps$, for $a\in A\setminus B$,
and $P(a)=a$, for $a\in B$. The {\em inverse image} of $P$,
denoted by $P^{-1} : B^* \to \pwr(A^*)$ with $\pwr(A^*)= 2^{A^*}$
being the power set of ${A^*}$, is defined as $P^{-1}(v)=\{w \in
A^* : P(w) = v\}$. These definitions can be extended to languages.
A generator $G$ is said to be partially observed if only a proper
subset of events $A_o\subset A$, called set of {\em observable
events\/}, is observed. The partial observation is described by
natural projection $O: A^* \to A_o^*$ defined as above with
$B=A_o$.

A {\em supervisor\/} based on partial observations for a controlled generator $(G,A_c,\Gamma)$
is a map $S: O(L(G)) \to \Gamma$. The {\em closed-loop system\/}
is denoted by $S/G$. The language generated by $S/G$, $L(S/G)$, is
defined recursively as (1) $\eps\in L(S/G)$ and, (2) for any $w\in
L(S/G)$ and $a \in A$, $wa \in L(S/G)$ if and only if $wa\in L(G)$
and $a \in S(O(w))$.

Given a specification (language) $K \subseteq L(G)$, the aim of
supervisory control under partial observations is to find a
supervisor $S$ such that $L(S/G)=K$. The existence condition for
such a supervisor is characterized by controllability and
observability defined as follows. $K$ is {\em controllable} with
respect to $L(G)$ and $A_{u}$ if ${K} A_{u} \cap L(G) \subseteq
{K}$ \cite{RW88}. $K$ is {\em observable} with respect to $L(G)$
and $A_o$ if $(\forall w, w' \in {K})~O(w)=O(w') \Rightarrow
(\forall a \in A) (wa \in {K} \wedge w'a \in L(G) \Rightarrow w'a
\in {K}) $ \cite{LinWonham2}.

It is proved in \cite{LinWonham2} that there exists a supervisor
that synthesizes $K$, that is, $L(S/G)=K$, if and only if $K$ is
controllable with respect to $L(G)$ and $A_{u}$ and observable
with respect to $L(G)$ and $A_{o}$.

For a generator $G$ and a projection $P$, $P(G)$ denotes the
minimal generator such that $L(P(G))=P(L(G))$. The reader is
referred to~\cite{CL08} for a construction of $P(G)$.
$P(G)$ is often called an {\em observer} of $G$.

In modular/coordination control, we consider local alphabets
(event sets) $A_i$, $A_j$, $A_\ell \subseteq A$, we use
$P^{i+j}_{\ell}$ to denote the projection from $(A_i\cup A_j)^*$
to $A_\ell^*$. If $A_i\cup A_j=A$, we simply write $P_\ell$.

The synchronous product of languages $L_i\subseteq A_i^*$,
$i=1,\dots ,n$, is defined as $\|_{i=1}^n L_i= \cap_{i=1}^n
P_i^{-1}(L_i) \subseteq A^*$, where $A = \cup_{i=1}^n A_i$ and
$P_i: A^*\to A_i^*$ are projections to local alphabets. For the
corresponding operation (also called synchronous product) in terms
of generators $G_i$, it is known that $L( \|_{i=1}^n G_i) =
\|_{i=1}^n L(G_i)$.

We now consider control of modular DES with a
global specification. The approach is based on the (relaxed)
coordination control of \cite{KMvS15}, where conditional
decomposability is used to bring the problem with global
specification to the problems of local specification.

Consider generators $G'_1$ and $G'_2$ over the alphabets $A'_1$
and $A'_2$, respectively. Let $G=G'_1 \| G'_2$ and $A=A'_1\cup
A'_2$. Given a prefix-closed specification $\overline{K}= K
\subseteq L(G)$, $K$ is {\em conditionally decomposable} with
respect to $A'_1$, $A'_2$, and $A'_k$ if $K = P_{1} (K) \parallel
P_{2} (K)$, where $P_{i}:A^*\to (A'_i \cup A'_k)^*$, $i=1,2$.

The following algorithm \cite{KMvS15} finds a coordinator $G'_k$
over $A'_k$ with $A'_1 \cap A'_2 \subseteq A'_k \subseteq A'_1\cup
A'_2=A$ and $P_k$ projection to $A_{k'}$ such that (1) $G'_k =
P_k(G'_1) \parallel P_k(G'_2)$, and (2) $K$ is conditionally
decomposable with respect to $A'_1$, $A'_2$, and $A'_k$. Note that
$G'_k = P_k(G'_1) \parallel P_k(G'_2)$ implies $G= G'_1 \parallel
G'_2 = G'_1 \parallel G'_2 \parallel G'_k$.

\begin{newalgorithm}(Construction of a Coordinator)
\label{algorithm}
Given $G'_1$ and $G'_2$ and $K \subseteq L(G)$, compute the event
set $A'_k$ and the coordinator $G'_k$ as follows.
\begin{enumerate}
\item
Let $A'_k = A'_1\cap A'_2$ be the set of all shared events of the
generators $G'_1$ and $G'_2$.
\item
Extend the alphabet $A'_k$ so that $K$ becomes conditional
decomposable with respect to $A'_1$, $A'_2$, and $A'_k$. (see
\cite{scl12} for a polynomial algorithm.)
\item
Define the coordinator $G'_k$ as $G'_k = P_k(G'_1) \parallel
P_k(G'_2)$.
\end{enumerate}
\end{newalgorithm}
It is well known that the computation of a projected generator
(observer) can be exponential in the worst case. However, it is
also known that if the projection satisfies the observer property
\cite{WW96}, then the projected generator is of the same order as
the original generator. Therefore, one might want to extend the
event set $A'_k$ further so that the projection $P_k:A^* \to
{A'}_k^*$ satisfies the observer property \cite{pcl08}.

\begin{figure}
\centering
  \includegraphics[scale=1.1]{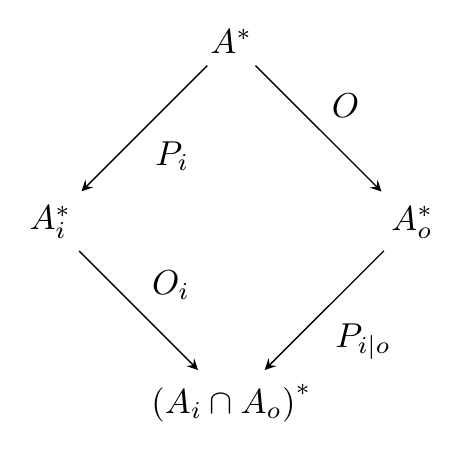}
\caption{Modular DES with partial observations}
\label{4projections}
\end{figure}

Denote $G_i=G'_i||G'_k$, $i=1,2$. Local supervisors $S_i$ operate
on $G_i$ over alphabets $A_{i}=A'_i \cup A'_k$. In this paper, we
assume that controllability and observability of events are
consistent over local supervisors, that is, $S_i$ can control
events in $A_{i,c}=A_i \cap A_c$ and observe events in
$A_{i,o}=A_i \cap A_o$. Let $A_{i,u}=A_i \setminus A_{i,c}$. Local
observation mapping is the projection $O_i: A_i^* \rightarrow
A_{i,o}^*$. The relation among $O$, $P_i$ and $O_i$ are shown in
Figure \ref{4projections}.
Local languages $P_{i} (K)\subseteq (A'_i \cup A'_k)^*, i=1,2$ are
then used as specifications for local supervisors in the
coordination control, given simply by supremal controllable
sublanguages or controllable and observable/normal sublanguages of
$P_i(K)$  with respect to $L(G_i)$, $A_{i,u}$ and $A_{i,o}$. In
the next section we will investigate algebraic conditions under
which this local (coordinated) control synthesis is as permissive
as the least restrictive monolithic synthesis.
\section{Comparison of Monolithic and Coordination Control}

If $K$ is not controllable and observable, then we would like to
find some sublanguage that is controllable and observable and we
would like to make such a sublanguage as large as possible. Let
$$
(K,L)^\Uparrow
$$
denote a sublanguage of $K$ which is either controllable (for $A_o=A$) or observable (for $A_c=A$) or both (in general) with respect to $L$. If $L$ is understood, then we use
$K^\Uparrow$ to denote $(K,L)^\Uparrow$. Note that $(K,L)^\Uparrow$ corresponds
to a closed-loop language, because controllability and observability characterize closed-loops. If $(K,L)^\Uparrow$ is
the supremal sublanguage (in several cases listed below it exists), 
we use $(K,L)^\uparrow$ to denote
$(K,L)^\Uparrow$.

We recall that controllability is preserved under language unions
and hence the supremal controllable sublanguage exists. On the
other hand, observability is not preserved under language unions.
Therefore, supremal observable sublanguages do not exist in
general. However, if all unobservable events are uncontrollable,
observability is equivalent to normality, which is defined as
follows. $K$ is {\em normal} with respect to $L$ and $A_o$ if
$O^{-1} O(K) \cap L \subseteq K$ \cite{LinWonham2}\footnote{Since
$K \subseteq O^{-1} O(K) \cap L$ is automatic, normality is
equivalent to $O^{-1} O(K) \cap L = K$.}. Normality is preserved
under language unions and hence the supremal normal sublanguage
exists.

More recently, relative observability has been introduced in
\cite{RO13, RO15} as a property weaker then normality but
preserved under language unions, which is defined as follows. Let
$K \subseteq C \subseteq L$. $K$ is {\em $C$-observable\/} with
respect to  $L$ and $A_o$ if $(\forall w \in {K})(\forall  w' \in
C)~O(w)=O(w') \Rightarrow (\forall a \in A) (wa \in {K}  \wedge w'a
\in L \Rightarrow w'a \in {K}) $. The supremal relatively
observable sublanguage exists.
\begin{example}
\label{example6}
Based on the above discussions, examples of $(K,L)^\uparrow$
include:
\begin{enumerate}
\item
the supremal controllable sublanguage of $K$, denoted by
$(K,L)^{\uparrow c}$ or $K^{\uparrow c}$,
\item
the supremal normal sublanguage of $K$, denoted by
$(K,L)^{\uparrow n}$ or $K^{\uparrow n}$,
\item
the supremal controllable and normal sublanguage of $K$, denoted
by $(K,L)^{\uparrow cn}$ or $K^{\uparrow cn}$,
\item
the  supremal $K$-observable sublanguage of $K$ with
respect to $L$, denoted by
$(K,L)^{\uparrow r}$ or $K^{\uparrow r}$,
\item
the  supremal $L$-observable sublanguage of $K$ with
respect to $L$, denoted by
$(K,L)^{\uparrow R}$ or $K^{\uparrow R}$.
\item
the supremal controllable and $K$-observable sublanguage of
$K$ with respect to $L$, denoted by $(K,L)^{\uparrow cr}$ or $K^{\uparrow cr}$.
\end{enumerate}
Examples of $(K,L)^\Uparrow$ include:
\begin{enumerate}
\item
a maximal observable sublanguage of $K$, denoted by
$(K,L)^{\Uparrow o}$ or $K^{\Uparrow o}$,
\item
a maximal controllable and observable sublanguage of $K$, denoted
by $(K,L)^{\Uparrow co}$ or $K^{\Uparrow co}$,
\item
a unique controllable and observable sublanguage of $K$ introduced
in \cite{HL94}, denoted by $(K,L)^{\Uparrow hl}$ or $K^{\Uparrow
hl}$.
\end{enumerate}
\end{example}
Let us compare the computational complexity and performance of
coordination control with those of monolithic control. It is well
known that the computational complexity of most algorithms for
partially observed DES is of the order
$O(2^{\|K\|})$, where $\|K\|$ denotes the number of states of the automaton 
generating the language $K$ with the minimal number of states.
To reduce the computational complexity, we use modular computation
of supervisors as follows. Assume that the specification language
is conditionally decomposable, that is, $K=P_{1}(K)\| P_{2}(K)$.
Denote
\begin{align*}
K_1 = P_{1}(K), \ \ \ & K_2 = P_{2}(K) \\
L_1 = L(G_1), \ \ \ & L_2 = L(G_2).
\end{align*}
Note $K=K_1 \| K_2$ and $L=L(G)=L_1 \| L_2$. We compute
$K_i^{\Uparrow}=(K_i, L_i) ^{\Uparrow}$, $i=1,2$ and use
coordination supervisors $S_i$ such that $L(S_i/G_i)= K_i
^{\Uparrow}$. We investigate the computational complexity and
performance of $S_1$ and $S_2$ vs the monolithic supervisor $S$ as
follows.

{\em Computational complexity of $S_1$ and $S_2$ vs $S$:}
Computational complexity of $S_1$ and $S_2$ is $O(2^{\|K_1\|}
+2^{\|K_2\|})$ and computational complexity of $S$ is
$O(2^{\|K\|})$. If $\|K\| = \|K_1\| \times \|K_2\|$ (by proper
construction of $G_k$), then $O(2^{\|K_1\|} +2^{\|K_2\|})$ is much
smaller than $O(2^{\|K\|})$.

{\em Performance of $S_1$ and $S_2$ vs $S$:} Since
\begin{align*}
L(S_1/G_1) = K_1^\Uparrow \ \ \ \mbox{and} \ \ \ L(S_2/G_2) =
K_2^\Uparrow ,
\end{align*}
the closed-loop system under coordination control is described by
\begin{align*}
L(S_1/G_1) \| L(S_2/G_2) = K_1^\Uparrow \| K_2^\Uparrow .
\end{align*}
For monolithic supervisor $S$,
\begin{align*}
L(S/G) = K^\Uparrow = (K_1 \| K_2)^\Uparrow .
\end{align*}
We first show the following obvious result for the sake of
completeness of the presentation. It states that both monolithic
supervisor and coordination supervisor ensure the safety, that is,
the language generated by the closed-loop system is within the
specification language $K$.
\begin{proposition} (Safety)
\label{proposition19}
Given $G_1$, $G_2$, and $K \subseteq L$. Let $K=K_1 \| K_2$ be
conditionally decomposable, then
\begin{align*}
& L(S/G) \subseteq K \\
& L(S_1/G_1) \| L(S_2/G_2) \subseteq K .
\end{align*}
\end{proposition}
{\bf Proof}
By the definition of the operation $(.)^\Uparrow$, $K^\Uparrow
\subseteq K$, $K_i^\Uparrow \subseteq K_i$, $i=1,2$. Hence
\begin{align*}
& L(S/G) = K^\Uparrow \subseteq K \\
& L(S_1/G_1) \| L(S_2/G_2) \\
& = K_1^\Uparrow \| K_2^\Uparrow \subseteq K_1 \| K_2 = K .
\end{align*}
\hfill$\square$

To compare $K_1^\Uparrow \| K_2^\Uparrow$ with $K^\Uparrow = (K_1
\| K_2)^\Uparrow$, we note that under some very general conditions
such as languages are prefix-closed, which we assume in this
paper,
$$
(K_1, L_1)^{\Uparrow} \| (K_2, L_2)^{\Uparrow} \subseteq (K,
L)^{\Uparrow}
$$
is always true \cite{KvS08}. Hence the key to the comparison is to
find conditions under which
$$
(K, L)^{\Uparrow} \subseteq (K_1, L_1)^{\Uparrow} \| (K_2,
L_2)^{\Uparrow}
$$
is true. To this end, let us first define some properties of the
operation $(.)^\Uparrow$ as follows.
\begin{definition} (Monotonicity)
\label{definition6}
The operation $(.)^\Uparrow$ is {\em monotonically increasing} if,
for all $K \subseteq L$ and $K' \subseteq L$,
\begin{align*}
K \subseteq K' \Rightarrow (K,L)^\Uparrow \subseteq
(K',L)^\Uparrow . 
\end{align*}
\end{definition}
\begin{definition} (Exchangeability)
\label{definition6}
Given $K_i \subseteq L_i \subseteq A_i^*$,
$(P_{i}^{-1}(K_i))^{\Uparrow}$ is {\em exchangeable} with respect
to $P_{i}$ if
\begin{align*}
(P_{i}^{-1}(K_i))^{\Uparrow} = P_{i}^{-1}(K_i^{\Uparrow}) .
\end{align*}
That is,
\begin{align*}
(P_{i}^{-1}(K_i), P_{i}^{-1}(L_i))^{\Uparrow} = P_{i}^{-1}((K_i,
L_i)^{\Uparrow}) . 
\end{align*}
\end{definition}
\begin{definition} (Distributivity)
\label{distributivity}
Given three languages $L$, $M$, and $K \subseteq L$,
$(K,L)^\Uparrow$ is {\em distributable} with respect to $M$ if
$$
(K \cap M,L \cap M)^\Uparrow = (K,L)^\Uparrow \cap M.
$$
\end{definition}
We now compare the performance of $S_1$ and $S_2$ vs $S$.
\begin{theorem} (Comparison)
\label{theorem10}
Given $K_i \subseteq L_i \subseteq A_i^*$, $i=1,2$, $K=K_1 ||
K_2$, and $L= L_1 || L_2$. Assume that (1) $(.)^\Uparrow$ is
monotonically increasing; (2) $(P_{i}^{-1}(K_i))^{\Uparrow}$ is
exchangeable with respect to $P_{i}$; and (3) $(P_{1}^{-1}(K_1),
P_{1}^{-1} (L_1))^{\Uparrow}$ is distributable with respect to
$P_{2}^{-1} (L_2)$ and $(P_{2}^{-1}(K_2),
P_{2}^{-1}(L_2))^{\Uparrow}$ is distributable with respect to
$P_{1}^{-1}(L_1)$. Then
\begin{align*}
K^{\Uparrow} \subseteq & K_1^{\Uparrow} \| K_2^{\Uparrow}.
\end{align*}
That is,
\begin{align*}
(K, L)^{\Uparrow} \subseteq (K_1, L_1)^{\Uparrow} \| (K_2,
L_2)^{\Uparrow} .
\end{align*}
\end{theorem}
{\bf Proof}
\begin{align*}
& (K,L)^{\Uparrow} =  (K_1 \| K_2, L)^{\Uparrow} \\
= & (K_1 \| K_2,L)^{\Uparrow} \cap (K_1 \| K_2,L)^{\Uparrow} \\
\subseteq & (K_1 \| L_2, L)^{\Uparrow} \cap (L_1 \| K_2,L)^{\Uparrow} \\
& \mbox{(by monotonicity)} \\
= & (P_{1}^{-1}(K_1) \cap P_{2}^{-1}(L_2), P_{1}^{-1}(L_1)
\cap P_{2}^{-1}(L_2))^{\Uparrow} \\
& \cap (P_{1}^{-1}(L_1) \cap P_{2}^{-1}(K_2), P_{1}^{-1}(L_1)
\cap P_{2}^{-1}(L_2))^{\Uparrow} \\
= & (P_{1}^{-1}(K_1), P_{1}^{-1}(L_1) )^{\Uparrow}
\cap P_{2}^{-1}(L_2) \\
& \cap (P_{2}^{-1}(K_2), P_{2}^{-1}(L_2))^{\Uparrow}  \cap
P_{1}^{-1}(L_1) \\
& \mbox{(by distributivity)} \\
= & (P_{1}^{-1}(K_1), P_{1}^{-1}(L_1) )^{\Uparrow} \cap
(P_{2}^{-1}(K_2), P_{2}^{-1}(L_2))^{\Uparrow} \\
= & P_{1}^{-1}((K_1,L_1)^{\Uparrow}) \cap
P_{2}^{-1}((K_2,L_2)^{\Uparrow}) \\
& \mbox{(by exchangeability)} \\
=& (K_1,L_1)^{\Uparrow} \| (K_2,L_2)^{\Uparrow}.
\end{align*}
\hfill$\square$
Theorem \ref{theorem10} provides three conditions on the closed-loops $(K,L)^{\Uparrow}$ that jointly guarantee that local closed-loops do not yield smaller result than the global closed-loop.   
If $(K,L)^\uparrow$ is the supremal sublanguages of $K$ that
satisfies the conditions in Proposition \ref{proposition12} below,
then monotonicity is satisfied automatically.
\begin{proposition}
\label{proposition12}
Assume that the operation $(.)^\uparrow$ satisfies the following
two conditions.

(1) $(K^\uparrow)^\uparrow = K^\uparrow$, and

(2) $(\forall K'' \subseteq K') (K'')^\uparrow = K'' \Rightarrow
K'' \subseteq (K')^\uparrow$. \\
Then $(.)^\uparrow$ is monotonically increasing, that is, for all
$M \subseteq L$ and $M' \subseteq L$,
$$
M \subseteq M' \Rightarrow M^\uparrow \subseteq (M')^\uparrow .
$$
\end{proposition}
{\bf Proof}
For $M \subseteq M'$,
$
M^\uparrow \subseteq M \subseteq M' .
$
Take $K''=M^\uparrow$ and $K'=M'$ in Condition (2). Since
$(M^\uparrow)^\uparrow = M^\uparrow$ (by Condition (1)),
Condition (2) gives
$$
M^\uparrow \subseteq (M')^\uparrow .
$$
\hfill$\square$

Since any supremal sublanguage of $K$ discussed in Example
\ref{example6} satisfies two conditions in Propositions
\ref{proposition12}, we have the following corollary.

\begin{corollary}
\label{corollary13}

Any supremal sublanguage of $K$ discussed in Example
\ref{example6} is monotonically increasing, that is, for all $K
\subseteq L$ and $K' \subseteq L$,
$
K \subseteq K' \Rightarrow K^\uparrow \subseteq (K')^\uparrow .
$
\end{corollary}

\section{Distributed Computation of Supremal Controllable Sublanguages}
\label{controllable_sublanguage}

Recall $K$ is controllable with respect to $L$ and $A_{u}$ if $K
A_{u} \cap L \subseteq K$. For local languages, $K_i$ is
controllable with respect to $L_i$ and $A_{i,u}$ if $K_i A_{i,u}
\cap L_i \subseteq K_i$. The supremal control sublanguages are
denote by $(K,L)^{\uparrow c}$ and $(K_i,L_i)^{\uparrow c}$
respectively. By Corollary \ref{corollary13}, $(.)^{\uparrow c}$
is monotonically increasing. We show that $(P_{i}^{-1}
(K_i))^{\uparrow c}$ is exchangeable in the following proposition.

\begin{proposition} (Exchangeability of $(.)^{\uparrow c}$)
\label{proposition11}

Given $K_i \subseteq L_i \subseteq A_i^*$,
$(P_{i}^{-1}(K_i))^{\uparrow c}$ is exchangeable with respect to
$P_{i}$, that is
\begin{align*}
(P_{i}^{-1}(K_i))^{\uparrow c} = P_{i}^{-1}(K_i^{\uparrow c}) .
\end{align*}
Or,
\begin{align*}
(P_{i}^{-1}(K_i), P_{i}^{-1}(L_i))^{\uparrow c} = P_{i}^{-1}((K_i,
L_i)^{\uparrow c}) .
\end{align*}

\end{proposition}
{\bf Proof}

Let the automata generating $K_i$ and $L_i$ be $H_i=(Q_H, A_i,
\delta_H, q_0)$ and $G_i=(Q, A_i, \delta, q_0)$ respectively, with
$H_i$ being a subautomaton of $G_i$, that is, $Q_H \subseteq Q$
and $\delta_H = \delta |_{Q_H}$. It is well-known that the
subautomaton
$$
H_i^{\uparrow c}=(Q_H^{\uparrow c}, A_i, \delta_H^{\uparrow c},
q_0)
$$
generating $K_i^{\uparrow c}$ can be obtained from $H_i$ by
removing the states from $Q_H$ that are co-accessible to $Q
\setminus Q_H$ via uncontrollable events, that is,
$$
Q_H^{\uparrow c} = Q_H \setminus \{q \in Q_H: (\exists s \in
A_{i,u}^*) \delta (q,s) \in Q \setminus Q_H \}
$$
and $\delta_H^{\uparrow c} = \delta_H|_{Q_H^{\uparrow c}}$.

The automata for $P_{i}^{-1}(K_i)$ and $P_{i}^{-1}(L_i)$ can be
obtained from $H_i$ and $G_i$ by adding self-loops of $A \setminus
A_i$ to all states. Denote the resulting automata by $\tH_i=(Q_H,
A, \tdelta _H, q_0)$ and $\tG_i=(Q, A, \tdelta, q_0)$
respectively. Since the only difference between $H_{i}$ and
$\tH_{i}$ ($G_{i}$ and $\tG_{i}$) is the self-loops of $A
\setminus A_i$, the states that are co-accessible to $Q \setminus
Q_H$ via uncontrollable events in $H_i$ and $\tH_i$ are same.
Hence,
$$
\tH_i^{\uparrow c}=(Q_H^{\uparrow c}, A, \tdelta_H^{\uparrow c},
q_0),
$$
where $\tdelta_H^{\uparrow c} = \tdelta_H|_{Q_H^{\uparrow c}}$.
Therefore,
\begin{align*}
& L(\tH_i^{\uparrow c} )= P_{i}^{-1}(L(H_i^{\uparrow c}))\\
\Rightarrow & (P_{i}^{-1}(K_i))^{\uparrow c} =
P_{i}^{-1}(K_i^{\uparrow c}).
\end{align*}
\hfill$\square$

The following proposition gives a sufficient condition for
distributivity of $(.)^{\uparrow c}$.

\begin{proposition} (Distributivity of $(.)^{\uparrow c}$)
\label{proposition15}

Given three languages $L$, $M$, and $K \subseteq L$, if $M$ is
controllable with respect to $L$, then  $(K,L)^{\uparrow c}$ is
distributable with respect to $M$, that is,
$$
(K \cap M)^{\uparrow c} = K^{\uparrow c} \cap M.
$$
Or,
$$
(K \cap M,L \cap M)^{\uparrow c} = (K,L)^{\uparrow c} \cap M.
$$
\end{proposition}
{\bf Proof}
\noindent ($\subseteq$:) Clearly,
$$
(K \cap M)^{\uparrow c} \subseteq K \cap M \subseteq M.
$$
To prove $(K \cap M)^{\uparrow c} \subseteq K^{\uparrow c}$, we
need to show the following. (1) $(K \cap M)^{\uparrow c} \subseteq
K$, which is clearly true. (2) $(K \cap M)^{\uparrow c}$ is
controllable with respect to $L$. To prove this, note that $(K
\cap M)^{\uparrow c}$ is controllable with respect to $L \cap M$.
On the other hand, $L$ is controllable with respect to itself,
$L$. By the assumption, $M$ is controllable with respect to $L$.
Since all languages are closed, $L \cap M$ is controllable with
respect to $L$. Hence, by the chain property of
controllability\footnote{If $M_1$ is controllable with respect to
$M_2$ and $M_2$ is controllable with respect to $M_3$, then $M_1$
is controllable with respect to $M_3$ \cite{scl11}.}, $(K \cap
M)^{\uparrow c}$ is controllable with respect to $L$.

\noindent ($\subseteq$:) To prove $K^{\uparrow c} \cap M \subseteq
(K \cap M)^{\uparrow c}$, we need to show the following. (1)
$K^{\uparrow c} \cap M \subseteq K \cap M$, which is clearly true.
(2) $K^{\uparrow c} \cap M$ is controllable with respect to $L
\cap M$. Indeed, this is true because
\begin{align*}
& (K^{\uparrow c} \cap M) A_u \cap L \cap M \\
\subseteq & (K^{\uparrow c}) A_u \cap M A_u \cap
L \cap M \\
\subseteq & (K^{\uparrow c}) A_u \cap
L \cap M \\
\subseteq & K^{\uparrow c} \cap M \\
& (\mbox{because }K^{\uparrow c} \mbox{ is controllable with
respect to } L)
\end{align*}
\hfill$\square$
From Proposition \ref{proposition15}, we conclude that in order to
use Theorem \ref{theorem10} for the supremal controllable
sublanguage $(.)^{\uparrow c}$, we need that (1)
$P_{1}^{-1} (L_1)$ is controllable with respect to
$P_{2}^{-1}(L_2)$ and (2) $P_{2}^{-1} (L_2)$ is controllable with
respect to $P_{1}^{-1}(L_1)$. These two
conditions are equivalent to global mutual controllability (GMC)
introduced in \cite{KvS07}. Let us recall the definition of GMC.

\begin{definition}
The modular plant languages $L_1$ and $L_2$ are {\em globally
mutually controllable} if  $\forall i,j \in \{1, 2\}, ~ i \not =
j$,
\begin{align*}
P_{i}^{-1} (L_i) A_{i,u} \cap P_{j}^{-1} (L_j) \subseteq
P_{i}^{-1} (L_i) .\\
\end{align*}
\end{definition}

Note that in the definition of GMC, local uncontrollable events
$A_{i,u}$ are used. While in the definition of $P_{i}^{-1} (L_i)$
being controllable with respect to $P_{j}^{-1}(L_j)$, global
uncontrollable events $A_{u}$ are used, that is,
$$
P_{i}^{-1} (L_i) A_{u} \cap P_{j}^{-1} (L_j) \subseteq P_{i}^{-1}
(L_i).
$$
However, it is proved in \cite{KvS07} that they are equivalent as
re-stated in the following Proposition.
\begin{proposition}
\label{proposition17}

The modular plant languages $L_1$ and $L_2$ are globally mutually
controllable if and only if (1) $P_{1}^{-1} (L_1)$ is controllable
with respect to $P_{2}^{-1}(L_2)$ and $A_u$ and (2) $P_{2}^{-1} (L_2)$ is
controllable with respect to $P_{1}^{-1}(L_1)$ and $A_u$.
\end{proposition}

Therefore, we have the following result for the supremal
controllable sublanguage $(.)^{\uparrow c}$.

\begin{theorem} (Comparison for $(.)^{\uparrow c}$)
\label{theorem13}

Given $K_i \subseteq L_i \subseteq A_i^*$, $i=1,2$, and $K=K_1 ||
K_2$, $L= L_1 || L_2$. If $L_1$ and $L_2$ are globally mutually
controllable, then
\begin{align*}
K^{\uparrow c} \subseteq & K_1^{\uparrow c} \| K_2^{\uparrow c}.
\end{align*}
That is,
\begin{align*}
(K, L)^{\uparrow c} \subseteq (K_1, L_1)^{\uparrow c} \| (K_2,
L_2)^{\uparrow c} .
\end{align*}

\end{theorem}
{\bf Proof}
By Corollary \ref{corollary13}, $(.)^{\uparrow c}$ is
monotonically increasing. By Proposition \ref{proposition11},
$(P_{i}^{-1} (K_i))^{\uparrow c}$ is exchangeable with respect to
$P_{i}$. By Propositions \ref{proposition15} and
\ref{proposition17}, $(P_{1}^{-1}(K_1), P_{1}^{-1}
(L_1))^{\uparrow c}$ is distributable with respect to $P_{2}^{-1}
(L_2)$ and $(P_{2}^{-1}(K_2), P_{2}^{-1}(L_2))^{\uparrow c}$ is
distributable with respect to $P_{1}^{-1}(L_1)$. Therefore, by
Theorem \ref{theorem10},
\begin{align*}
K^{\uparrow c} \subseteq & K_1^{\uparrow c} \| K_2^{\uparrow c}.
\end{align*}
\hfill$\square$

Note that the above proof is based on arguments that do not depend
on the particular property (in this case, controllability).
Therefore, we can extend this result for distributed computation
of languages arising in supervisory control with partial
observations in the next two sections.

In the literature, there exists a well known concept of mutual
controllability (MC) \cite{Lee} that also ensures $(K_1,
L_1)^{\uparrow c} \| (K_2, L_2)^{\uparrow c} =(K, L)^{\uparrow
c}$. In the rest of this section we compare GMC with MC. Let us
recall the definition of MC.

\begin{definition}
The modular plant languages $L_1$ and $L_2$ are {\em mutually
controllable} if for all $i,j \in \{1, 2\}, ~ i \not = j$,
\begin{align*}
L_i (A_{i,u} \cap A_j) \cap P_{i} (P_{j}^{-1} (L_j)) \subseteq L_i
. \\
\end{align*}
\end{definition}

To compare GMC with MC, we show that MC is equivalent to the
following weakly globally mutual controllability (WGMC).

\begin{definition}
The modular plant languages $L_1$ and $L_2$ are {\em weakly
globally mutually controllable} if all $i,j \in \{1, 2\}, ~ i \not
= j$,
\begin{align*}
P_{i}^{-1} (L_i) (A_{i,u} \cap A_{j}) \cap P_{j}^{-1} (L_j)
\subseteq P_{i}^{-1} (L_i) .
\end{align*}
\end{definition}

The following proposition states the relation between WGMC and MC.

\begin{proposition}
\label{proposition21}

Weak global mutual controllability is equivalent to  mutual
controllability.
\end{proposition}
{\bf Proof}

First we show that MC implies WGMC. Let MC be true, that is, for
all $i,j \in \{1, 2\}, ~ i \not = j$,
\begin{align*}
 L_j(A_{ju}\cap A_i)\cap P_j(P_i)^{-1}(L_i)
  \subseteq L_j .
\end{align*}
By applying the inverse projection $P_j^{-1}$, we get, by
monotonicity of $P_j^{-1}$, that
\begin{align*}
P_j^{-1}[L_j(A_{ju}\cap A_i)\cap P_j(P_i)^{-1}(L_i)] \subseteq
P_j^{-1} L_j .
\end{align*}
Because inverse projection preserves catenation and intersections,
we obtain
\begin{align*}
P_j^{-1}(L_j) P_j^{-1}(A_{ju}\cap A_i)\cap P_j^{-1}
P_j(P_i)^{-1}(L_i)] \subseteq P_j^{-1} L_j .
\end{align*}
Since $A_{ju}\cap A_i \subseteq P_j^{-1}(A_{ju}\cap A_i)$ and
$(P_i)^{-1}(L_i) \subseteq P_j^{-1} P_j(P_i)^{-1}(L_i)$ we have
\begin{align*}
P_j^{-1}(L_j)(A_{ju}\cap A_i) \cap P_i^{-1}(L_i) \subseteq
P_j^{-1}(L_j) ,
\end{align*}
Hence, WGMC is true.

The inverse implication can be proved as follows. Let WGMC be
true, that is, for all $i,j \in \{1, 2\}, ~ i \not = j$,
\begin{align*}
P_j^{-1}(L_j)(A_{ju}\cap A_i) \cap P_i^{-1}(L_i) \subseteq
P_j^{-1}(L_j) .
\end{align*}
Note that
 \Lfteqn
\label{in} & P_j^{-1}(L_j)(A_{ju}\cap A_i) \cap
P_i^{-1}(L_i)\\
\subseteq & P_j^{-1}(L_j)P_j^{-1}(A_{ju}\cap A_i) \cap
P_i^{-1}(L_i)\\
= &L_j(A_{ju}\cap A_i)  \| L_i.
 \Ndeqn
WGMC implies that
\begin{align*}
   P_j[ P_j^{-1}(L_j)(A_{ju}\cap A_i) \cap P_i^{-1}(L_i)]
  \subseteq  P_jP_j^{-1}(L_j)=L_j .
\end{align*}
We first show that $~ \forall i,j \in \{1,2 \}, ~ i \not = j$,
$P_j[L_j(A_{ju}\cap A_i)  \| L_i] \subseteq  L_j$ already implies
MC.

Indeed, by conditional independence property we get that the last
statement implies $P_j[L_j(A_{ju}\cap A_i)  \| L_i]=
P_j[L_j(A_{ju}\cap A_i)] \|  P_j(L_i)= L_j(A_{ju}\cap A_i)\cap
P_i^{-1}P_j(L_i) =
 L_j(A_{ju}\cap A_i)\cap  P_jP_i^{-1}(L_i)\subseteq  L_j$,
 which is MC.

It remains to show that the obviously strict
inclusion of (\ref{in}) becomes equality when $P_j$ is applied to
both sides. In fact,
\begin{align*}
& P_j[ P_j^{-1}(L_j)(A_{ju}\cap A_i) \cap P_i^{-1}(L_i)] \\
= & P_j[ P_j^{-1}(L_j)P_j^{-1}((A_{ju}\cap A_i) \cap
P_i^{-1}(L_i)].
\end{align*}

The nontrivial inclusion is proven below. Let $s\in P_j[
P_j^{-1}(L_j)P_j^{-1}((A_{ju}\cap A_i) \cap P_i^{-1}(L_i)]$. Then
$s=P_j(t)$ for some $t\in P_j^{-1}(L_j)P_j^{-1}((A_{ju}\cap A_i)
\cap P_i^{-1}(L_i)$. Since $P_j\circ P_j^{-1}$ is identity, this
simply means that $s=s'u$ for $s'\in L_j$ and $u\in A_{ju}\cap
A_i$.

But then for any $t'\in P_j^{-1}(s')$ we have that
$su=P_j(t'u)=P_j(t')P_j(u)=P_j(t')u $, hence $s\in P_j[
P_j^{-1}(L_j)P_j^{-1}((A_{ju}\cap A_i) \cap P_i^{-1}(L_i)]$ as
well. Therefore, we get from WGMC by applying $P_j$ to both sides
\begin{align*}
   P_j[ P_j^{-1}(L_j)(A_{ju}\cap A_i) \cap P_i^{-1}(L_i)]
  \subseteq  P_jP_j^{-1}(L_j)=L_j ,
\end{align*}
also that $P_j[ P_j^{-1}(L_j)(A_{ju}\cap A_i) \cap P_i^{-1}(L_i)]
\subseteq P_jP_j^{-1}(L_j)=L_j, ~ \forall i,j \in \{1, 2\}, ~ i
\not = j,$ from which we have already derived above MC.

\hfill$\square$

Comparing the definitions of GMC and WGMC, it is clear that GMC is
stronger than WGMC, that is, GMC implies WGMC. We also have a
counter example that shows GMC is strictly stronger than WGMC. By
Proposition \ref{proposition21}, GMC is strictly stronger than MC.
The advantage of using GMC is that it is easier to check GMC than
to check MC, because it uses only computationally cheap inverse
projections, while MC uses both projections and inverse
projections.

\section{Distributed Computation of Supremal Normal Sublanguages}
\label{normal_sublanguage}
In this section we show how structural conditions proposed in
Section 3 can be used for distributed computation of supremal
normal sublanguages. We assume that at least one local supervisor
does not have complete observations. 

The local alphabets admit a partition into
locally observable and locally unobservable event sets as
specified in Section II. We recall that observability of events are
consistent over local supervisors, which can also be stated as 
$A_{1,o} \cap A_2 = A_{1} \cap A_{2,o} $.

Recall $K$ is normal with respect to $L$ and $A_o$ if $O^{-1} O(K)
\cap L \subseteq K$. For local languages, $K_i$ is normal with
respect to $L_i$ and $A_{i,o}$ if $O_i^{-1} O_i(K_i) \cap L_i
\subseteq K_i$. The supremal normal sublanguages are denote by
$(K,L)^{\uparrow n}$ and $(K_i,L_i)^{\uparrow n}$ respectively. By
Corollary \ref{corollary13}, $(.)^{\uparrow n}$ is monotonically
increasing. We show that $(P_{i}^{-1} (K_i))^{\uparrow n}$ is
exchangeable in the following proposition.

\begin{proposition} (Exchangeability of $(.)^{\uparrow n}$)
\label{proposition22}
Given $K_i \subseteq L_i \subseteq A_i^*$,
$(P_{i}^{-1}(K_i))^{\uparrow n}$ is exchangeable with respect to
$P_{i}$, that is
\begin{align*}
(P_{i}^{-1}(K_i))^{\uparrow n} = P_{i}^{-1}(K_i^{\uparrow n}) .
\end{align*}
Or,
\begin{align*}
(P_{i}^{-1}(K_i), P_{i}^{-1}(L_i))^{\uparrow n} = P_{i}^{-1}((K_i,
L_i)^{\uparrow n}) .
\end{align*}
\end{proposition}
{\bf Proof}
Let the automata generating $K_i$ and $L_i$ be $H_i=(Q_H, A_i,
\delta_H, q_0)$ and $G_i=(Q, A_i, \delta, q_0)$ respectively, with
$H_i$ being a subautomaton of $G_i$. It can be proven from results
in \cite{Brandt} that the subautomaton
$$
H_i^{\uparrow n}=(Q_H^{\uparrow n}, A_i, \delta_H^{\uparrow n},
q_0)
$$
generating $K_i^{\uparrow n}$ can be obtained from $H_i$ by
removing the states from $Q_H$ that are indistinguishable from
states in $Q \setminus Q_H$, that is,
\begin{align*}
Q_H^{\uparrow n} = & Q_H \setminus \{q \in Q_H: (\exists s, s' \in
L_i) O_i(s)=Q_i(s') \\
& \wedge \delta (q_0,s)=q \wedge \delta (q_0,s') \in Q \setminus
Q_H \}
\end{align*}
and $\delta_H^{\uparrow n} = \delta_H|_{Q_H^{\uparrow n}}$.

The automata for $P_{i}^{-1}(K_i)$ and $P_{i}^{-1}(L_i)$ can be
obtained from $H_i$ and $G_i$ by adding self-loops of $A \setminus
A_i$ to all states. Denote the resulting automata by $\tH_i=(Q_H,
A, \tdelta _H, q_0)$ and $\tG_i=(Q, A, \tdelta, q_0)$
respectively. Since the only difference between $H_{i}$ and
$\tH_{i}$ ($G_{i}$ and $\tG_{i}$) is the self-loops of $A
\setminus A_i$, the states that are are indistinguishable from
states in $Q \setminus Q_H$ in $H_i$ and $\tH_i$ are same, because
$(\forall s, s' \in P_i ^{-1}(L_i)) O(s)=Q(s') \Leftrightarrow
O_i(P_i(s))=O_i(P_i(s'))$. Hence,
$$
\tH_i^{\uparrow n}=(Q_H^{\uparrow n}, A, \tdelta_H^{\uparrow n},
q_0),
$$
where $\tdelta_H^{\uparrow n} = \tdelta_H|_{Q_H^{\uparrow n}}$.
Therefore,
\begin{align*}
& L(\tH_i^{\uparrow n} )= P_{i}^{-1}(L(H_i^{\uparrow n}))\\
\Rightarrow & (P_{i}^{-1}(K_i))^{\uparrow n} =
P_{i}^{-1}(K_i^{\uparrow n}).
\end{align*}
\hfill$\square$
The following proposition gives a sufficient condition for
distributivity of $(.)^{\uparrow n}$.
\begin{proposition} (Distributivity of $(.)^{\uparrow n}$)
\label{proposition23}
Given three languages $L$, $M$, and $K \subseteq L$, if $M$ is
normal with respect to $L$, then  $(K,L)^{\uparrow n}$ is
distributable with respect to $M$, that is,
$$
(K \cap M)^{\uparrow n} = K^{\uparrow n} \cap M.
$$
Or,
$$
(K \cap M,L \cap M)^{\uparrow n} = (K,L)^{\uparrow n} \cap M.
$$
\end{proposition}
{\bf Proof}
\noindent ($\subseteq$:) Clearly,
$$
(K \cap M)^{\uparrow n} \subseteq K \cap M \subseteq M.
$$
To prove $(K \cap M)^{\uparrow n} \subseteq K^{\uparrow n}$, we
need to show the following. (1) $(K \cap M)^{\uparrow n} \subseteq
K$, which is clearly true. (2) $(K \cap M)^{\uparrow n}$ is normal
with respect to $L$. To prove this, note that $(K \cap
M)^{\uparrow n}$ is normal with respect to $L \cap M$. Let us show
that $L \cap M$ is normal with respect to $L$ as follows.
\begin{align*}
& O^{-1}O(L \cap M) \cap L  \\
\subseteq & O^{-1}O(L) \cap O^{-1}O(M) \cap L\\
= & O^{-1}O(M) \cap L \subseteq  M \cap L \\
& (\mbox{because }M \mbox{ is normal with respect to } L)
\end{align*}
By the chain property of normality\footnote{If $M_1$ is normal
with respect to $M_2$ and $M_2$ is normal with respect to $M_3$,
then $M_1$ is normal with respect to $M_3$ \cite{scl11}.}, $(K
\cap M)^{\uparrow n}$ is normal with respect to $L$.

\noindent ($\subseteq$:) To prove $K^{\uparrow n} \cap M \subseteq
(K \cap M)^{\uparrow n}$, we need to show the following. (1)
$K^{\uparrow n} \cap M \subseteq K \cap M$, which is clearly true.
(2) $K^{\uparrow n} \cap M$ is normal with respect to $L \cap M$.
Indeed, this is true because
\begin{align*}
& O^{-1}O(K^{\uparrow n} \cap M) \cap L \cap M \\
\subseteq & O^{-1}O(K^{\uparrow n}) \cap O^{-1}O(M) \cap
L \cap M \\
= & O^{-1}O(K^{\uparrow n}) \cap L \cap M \\
\subseteq & K^{\uparrow n} \cap M \\
& (\mbox{because }K^{\uparrow n} \mbox{ is normal with respect to
} L)
\end{align*}
\hfill$\square$
From Proposition \ref{proposition23}, we conclude that in order to
use Theorem \ref{theorem10} for the supremal normal sublanguage
$(.)^{\uparrow n}$, it is required that (1) $P_{1}^{-1} (L_1)$ is
normal with respect to $P_{2}^{-1}(L_2)$ and (2) $P_{2}^{-1}
(L_2)$ is normal with respect to $P_{1}^{-1}(L_1)$. We call this
requirement globally mutual normality (GMN), which is defined as
follows.
\begin{definition}
The modular plant languages $L_1$ and $L_2$ are {\em globally
mutually normal} if for all $i,j \in \{1, 2\}, ~ i \not = j$,
$$
O^{-1}O (P_{i}^{-1} (L_i)) \cap P_{j}^{-1} (L_j) \subseteq
P_{i}^{-1} (L_i) .
$$
\end{definition}
We then have the following result for the supremal normal
sublanguage $(.)^{\uparrow n}$.
\begin{theorem} (Comparison for $(.)^{\uparrow n}$)
\label{theorem25}
Given $K_i \subseteq L_i \subseteq A_i^*$, $i=1,2$, and $K=K_1 ||
K_2$, $L= L_1 || L_2$. If $L_1$ and $L_2$ are globally mutually
normal, then
\begin{align*}
K^{\uparrow n} \subseteq & K_1^{\uparrow n} \| K_2^{\uparrow n}.
\end{align*}
That is,
\begin{align*}
(K, L)^{\uparrow n} \subseteq (K_1, L_1)^{\uparrow n} \| (K_2,
L_2)^{\uparrow n} .
\end{align*}
\end{theorem}
{\bf Proof}
By Corollary \ref{corollary13}, $(.)^{\uparrow n}$ is
monotonically increasing. By Proposition \ref{proposition22},
$(P_{i}^{-1} (K_i))^{\uparrow n}$ is exchangeable with respect to
$P_{i}$. Since $L_1$ and $L_2$ are globally mutually normal, by
Propositions \ref{proposition23}, $(P_{1}^{-1}(K_1), P_{1}^{-1}
(L_1))^{\uparrow n}$ is distributable with respect to $P_{2}^{-1}
(L_2)$ and $(P_{2}^{-1}(K_2), P_{2}^{-1}(L_2))^{\uparrow n}$ is
distributable with respect to $P_{1}^{-1}(L_1)$. Thus, by
Theorem \ref{theorem10},
\begin{align*}
K^{\uparrow n} \subseteq & K_1^{\uparrow n} \| K_2^{\uparrow n}.
\end{align*}
\hfill$\square$
To compare the above result with existing results in the
literature, we show that GMN is equivalent to mutual normality
(MN) introduced in \cite{KvS07}. Let us recall the definition of
MN and compare it with GMN.
\begin{definition}
The modular plant languages $L_1$ and $L_2$ are {\em mutually
normal} if for all $i,j \in \{1, 2\}, ~ i \not = j$,
\begin{align*}
O_i^{-1}O_i(L_i)\cap P_i(P_j^{-1}(L_j)) \subseteq L_i.
\end{align*}
\end{definition}
\begin{proposition}
\label{proposition25}
The modular plant languages $L_1$ and $L_2$ are globally mutually
normal if and only if they are mutually normal.
\end{proposition}
{\bf Proof}
We first prove
 \Lfteqn
 \label{Equation2}
O^{-1}O (P_{i}^{-1} (L_i)) = P_{i}^{-1} (O_i^{-1}O_i (L_i)).
 \Ndeqn
Let $G_i=(Q, A_i, \delta, q_0)$ be the automata generating $L_i$
and $G_{i,obs}=(X, A_{i,o}, \xi, x_0)$ be the observer generating
$O_i (L_i)$. By adding self-loops of $A \setminus A_{i,o}$ to
$G_{i,obs}$, we obtain the automaton $\tG_{i,obs}=(X, A, \txi,
x_0)$. Clearly $L(\tG_{i,obs})=P_{i}^{-1} (O_i^{-1}O_i (L_i))$. On
the other hand, adding self-loops of $A \setminus A_{i}$ and then
building observer for $O$ is same as building observer and then
adding self-loops. Hence, $L(\tG_{i,obs})=O^{-1}O (P_{i}^{-1}
(L_i))$ as well.

Let us now prove that MN implies GMN as follows.
\begin{align*}
& O_i^{-1}O_i(L_i)\cap P_i(P_j^{-1}(L_j)) \subseteq L_i \\
\Rightarrow & P_i^{-1} (O_i^{-1}O_i(L_i)\cap P_i(P_j^{-1}(L_j)))
\subseteq P_i^{-1} (L_i) \\
\Leftrightarrow & P_i^{-1} (O_i^{-1}O_i(L_i)) \cap P_i^{-1}
(P_i(P_j^{-1}(L_j))) \subseteq P_i^{-1} (L_i) \\
& \mbox{(because $P_i^{-1}$ preserves intersections)}\\
\Leftrightarrow & P_i^{-1} (O_i^{-1}O_i(L_i)) \cap P_j^{-1}(L_j)
\subseteq P_i^{-1} (L_i) \\
\Leftrightarrow & O^{-1}O (P_{i}^{-1} (L_i)) \cap P_j^{-1}(L_j)
\subseteq P_i^{-1} (L_i) \\
& (\mbox{by Equation (\ref{Equation2})})
\end{align*}
To prove that GMN implies MN, we do the following.
\begin{align*}
& O^{-1}O (P_{i}^{-1} (L_i)) \cap P_j^{-1}(L_j) \subseteq P_i^{-1}
(L_i) \\
\Leftrightarrow & P_i^{-1} (O_i^{-1}O_i(L_i)) \cap P_j^{-1}(L_j)
\subseteq P_i^{-1} (L_i) \\
& (\mbox{by Equation (\ref{Equation2}})) \\
\Leftrightarrow & O_i^{-1}O_i(L_i) \| L_j \subseteq P_i^{-1}
(L_i) \\
\Rightarrow & P_i(O_i^{-1}O_i(L_i) \| L_j) \subseteq P_i
(P_i^{-1}(L_i))= L_i
\end{align*}
It is proved in \cite{phdthesis} that, if the event set which
$P_i$ projects to ($A_i$ in the current case) contains the shared
events of the shuffle $\|$ ($A_i \cap A_j$ in the current case),
then $P_i$ can be distributed over $\|$. Hence
\begin{align*}
& P_i(O_i^{-1}O_i(L_i) \| L_j) \subseteq L_i \\
\Leftrightarrow & P_i(O_i^{-1}O_i(L_i)) \| P_i(L_j) \subseteq L_i \\
\Leftrightarrow & O_i^{-1}O_i(L_i) \| P_i(L_j) \subseteq L_i \\
\Leftrightarrow & O_i^{-1}O_i(L_i)\cap P_i(P_j^{-1}(L_j))
\subseteq L_i.
\end{align*}
\hfill$\square$
We conclude that mutual normality and global mutual normality are
equivalent. 

\section{Distributed Computation of Supremal Relatively Observable Sublanguages}
\label{ro_sublanguage}
In this section, we investigate relative observability. We first
take $C=K$. Clearly, $K$ is observable if and only if $K$ is
$K$-observable. However, we emphasize that relative observability
is preserved by language unions. Hence, for any two sublanguages
$K_1$ and $K_2$ of $K$, if both $K_1$ and $K_2$ are
$K$-observable, then $K_1 \cup K_2$ is also $K$-observable.
Therefore, the supremal $K$-observable sublanguage of $K$ with
respect to $L$ and $A_o$ exists and is denoted by $(K,L)^{\uparrow
r}$. For local languages $K_i$ and $L_i$, the supremal
$K_i$-observable sublanguage of $K_i$ with respect to $L_i$ and
$A_{i,o}$ is denoted by $(K_i,L_i)^{\uparrow r}$. In the
literature, there exist several different algorithms for
computation of supremal relatively observable sublanguages, see
\cite{RO15}, for example.

By Corollary \ref{corollary13}, $(.)^{\uparrow r}$ is
monotonically increasing. We show that $(P_{i}^{-1}
(K_i))^{\uparrow r}$ is exchangeable.
\begin{proposition} (Exchangeability of $(.)^{\uparrow r}$)
\label{proposition28}
Given $K_i \subseteq L_i \subseteq A_i^*$, we have
\begin{align*}
(P_{i}^{-1}(K_i))^{\uparrow r} = P_{i}^{-1}(K_i^{\uparrow r}) .
\end{align*}
Or,
\begin{align*}
(P_{i}^{-1}(K_i), P_{i}^{-1}(L_i))^{\uparrow r} = P_{i}^{-1}((K_i,
L_i)^{\uparrow r}).
\end{align*}
\end{proposition}
{\bf Proof}
Let the automata generating $K_i$ and $L_i$ be $H_i=(Q_H, A_i,
\delta_H, q_0)$ and $G_i=(Q, A_i, \delta, q_0)$ respectively, with
$H_i$ being a subautomaton of $G_i$. Based on these two automata,
an algorithm is given in \cite{RO15} to construct 
$$
H_i^{\uparrow r}=(Q_H^{\uparrow r}, A_i, \delta_H^{\uparrow r},
q_0)
$$
that generates $K_i^{\uparrow r}$. The algorithm can be viewed as
a refinement of the algorithm that constructs an automaton
$H_i^{\uparrow n}$ generating $K_i^{\uparrow n}$ outlined in the
proof of Proposition \ref{proposition22}. It is clear from the
algorithm that the addition of self-loops of events not in $A_i$ to all states,
does not change the result.

The automata for $P_{i}^{-1}(K_i)$ and $P_{i}^{-1}(L_i)$ can be
obtained from $H_i$ and $G_i$ by adding self-loops of $A \setminus
A_i$ to all states. Denote the resulting automata by $\tH_i=(Q_H,
A, \tdelta _H, q_0)$ and $\tG_i=(Q, A, \tdelta, q_0)$
respectively. Since the only difference between $H_{i}$ and
$\tH_{i}$ ($G_{i}$ and $\tG_{i}$) is the self-loops of $A
\setminus A_i$, the automaton generating $(P_{i}^{-1}(K_i))
^{\uparrow r}$ can be obtained from $H_i^{\uparrow r}$ by adding
the self-loops of $A \setminus A_i$. Denote the resulting
automaton by
$$
\tH_i^{\uparrow r}=(Q_H^{\uparrow r}, A, \tdelta_H^{\uparrow r},
q_0). \mbox{ Then}
$$
\begin{align*}
& L(\tH_i^{\uparrow r} )= P_{i}^{-1}(L(H_i^{\uparrow r}))\\
\Rightarrow & (P_{i}^{-1}(K_i))^{\uparrow r} =
P_{i}^{-1}(K_i^{\uparrow r}).
\end{align*}
\hfill$\square$
We now prove the following proposition.
\begin{proposition} (Distributivity of $(.)^{\uparrow r}$)
\label{proposition30r}
Given three languages $L$, $M$, and $K
\subseteq L$, if $M$ is $K$-observable with respect to
$L$\footnote{In the definition of relative observability
\cite{RO15}, it is assumed that $M \subseteq K \subseteq L$. This
is needed in order to show $M$ is $K$-observable with respect to
$L$ $\Rightarrow$ $M$ is observable with respect to $L$. This
implication is not needed for this proposition. Therefore, we
relax the assumption of $M \subseteq K \subseteq L$.}, then
$(K,L)^{\uparrow r}$ is distributable with respect to $M$,i.e.
$$
(K \cap M)^{\uparrow r} = K^{\uparrow r} \cap M.
$$
Or,
$$
(K \cap M,L \cap M)^{\uparrow r} = (K,L)^{\uparrow r} \cap M.
$$
\end{proposition}
{\bf Proof}
It is not difficult to see that $M$ is $K$-observable with respect
to $L$ if and only if
 \Lfteqn
 \label{Equation3}
(\forall w \in A^*)(\forall w' \in O^{-1}O(w))(\forall a \in A) \\
wa \in M \wedge w' \in K \wedge w'a \in L \Rightarrow w'a \in M.
 \Ndeqn
Similarly, since $(K \cap M)^{\uparrow r}$ is $(K \cap
M)$-observable with respect to $(L \cap M)$,
 \Lfteqn
 \label{Equation4}
(\forall w \in A^*)(\forall w' \in O^{-1}O(w))(\forall a \in A) \\
wa \in (K \cap M)^{\uparrow r} \wedge w' \in (K \cap M) \\
\wedge w'a \in (L \cap M) \Rightarrow w'a \in (K \cap M)^{\uparrow
r}.
 \Ndeqn
Since $K^{\uparrow r}$ is $K$-observable with respect to $L$,
 \Lfteqn
 \label{Equation5}
(\forall w \in A^*)(\forall w' \in O^{-1}O(w))(\forall a \in A) \\
wa \in K^{\uparrow r} \wedge w' \in K \wedge w'a \in L \Rightarrow
w'a \in K^{\uparrow r}.
 \Ndeqn
We now prove the result. \newline
\noindent ($\subseteq$:) Clearly,
$$
(K \cap M)^{\uparrow r} \subseteq K \cap M \subseteq M.
$$
To prove $(K \cap M)^{\uparrow r} \subseteq K^{\uparrow r}$, we
need to show that \\
(1) $(K \cap M)^{\uparrow r} \subseteq K$ (obvious) and \\
(2) $(K \cap M)^{\uparrow r}$ is $K$-observable with respect to
$L$, that is,
\begin{align*}
&(\forall w \in A^*)(\forall w' \in O^{-1}O(w))(\forall a \in A) \\
&wa \in (K \cap M)^{\uparrow r} \wedge w' \in K
\wedge w'a \in L \\
&\Rightarrow w'a \in (K \cap M)^{\uparrow r}.
\end{align*}
This is true because
\begin{align*}
&wa \in (K \cap M)^{\uparrow r} \wedge w' \in K
\wedge w'a \in L \\
\Rightarrow &wa \in (K \cap M)^{\uparrow r} \wedge w' \in K
\wedge w'a \in L \wedge wa \in M \\
\Rightarrow &wa \in (K \cap M)^{\uparrow r} \wedge w' \in K
\wedge w'a \in L \wedge w'a \in M \\
& (\mbox{by Equation (\ref{Equation3})}) \\
\Rightarrow &wa \in (K \cap M)^{\uparrow r} \wedge w' \in (K \cap
M) \wedge w'a \in (L \cap M) \\
\Rightarrow & w'a \in (K \cap M)^{\uparrow r} \; (\mbox{by Equation (\ref{Equation4})}).
\end{align*}
\noindent ($\supseteq$:) To prove $K^{\uparrow r} \cap M \subseteq
(K \cap M)^{\uparrow r}$, we need to show the following. (1)
$K^{\uparrow r} \cap M \subseteq K \cap M$, which is clearly true.
(2) $K^{\uparrow r} \cap M$ is $(K \cap M)$-observable with
respect to $(L \cap M)$, that is,
\begin{align*}
&(\forall w \in A^*)(\forall w' \in O^{-1}O(w))(\forall a \in A) \\
&wa \in (K^{\uparrow r} \cap M) \wedge w' \in (K \cap M) \\
&\wedge w'a \in (L \cap M) \Rightarrow w'a \in (K^{\uparrow r}
\cap M).
\end{align*}
This is true because
\begin{align*}
& wa \in (K^{\uparrow r} \cap M) \wedge w' \in (K \cap M) \wedge
w'a \in (L \cap M) \\
\Rightarrow & (wa \in K^{\uparrow r} \wedge w' \in K \wedge w'a
\in L) \wedge w'a \in M \\
\Rightarrow & wa' \in K^{\uparrow r} \wedge w'a \in M \; (\mbox{by Equation (\ref{Equation5})}) \\
\Rightarrow & wa' \in (K^{\uparrow r} \cap M) .
\end{align*}
\hfill$\square$
From Proposition \ref{proposition30r}, we conclude that in order
to use Theorem \ref{theorem10} for the supremal relatively
observable sublanguage $(.)^{\uparrow r}$, it is required that (1)
$P_{1}^{-1} (L_1)$ is $P_{1}^{-1} (K_2)$-observable with respect
to $P_{2}^{-1}(L_2)$ and (2) $P_{2}^{-1} (L_2)$ is $P_{1}^{-1}
(K_1)$-observable with respect to $P_{1}^{-1}(L_1)$. 
\begin{definition}
The modular plant languages $L_1$ and $L_2$ are {\em globally
mutually $K$-observable} if for all $i,j \in \{1, 2\}, ~ i \not = j$,
\begin{align*}
&(\forall w \in A^*)(\forall w' \in O^{-1}O(w))(\forall a \in A) \\
& wa \in P_{i}^{-1} (L_i)) \wedge w' \in P_{j}^{-1} (K_j) \wedge
w'a \in P_{j}^{-1} (L_j)) \\
& \Rightarrow w'a \in P_{i}^{-1}
(L_i)). 
\end{align*}
\end{definition}
We then have the following result.
\begin{theorem} (Comparison for $(.)^{\uparrow r}$)
\label{theorem25}
Let $K_i \subseteq L_i \subseteq A_i^*$, $i=1,2$, $K=K_1 ||
K_2$, and  $L= L_1 || L_2$. If $L_1$ and $L_2$ are globally mutually
$K$-observable then
\begin{align*}
K^{\uparrow r} \subseteq & K_1^{\uparrow r} \| K_2^{\uparrow r}.
\end{align*}
That is,
\begin{align*}
(K, L)^{\uparrow r} \subseteq (K_1, L_1)^{\uparrow r} \| (K_2,
L_2)^{\uparrow r}.
\end{align*}
\end{theorem}
{\bf Proof}
By Corollary \ref{corollary13}, $(.)^{\uparrow r}$ is
monotonically increasing. By Proposition \ref{proposition28},
$(P_{i}^{-1} (K_i))^{\uparrow r}$ is exchangeable with respect to
$P_{i}$. Since $L_1$ and $L_2$ are globally mutually $K$-observable,
by Proposition \ref{proposition30r}, $(P_{1}^{-1}(K_1), P_{1}^{-1}
(L_1))^{\uparrow r}$ is distributable with respect to $P_{2}^{-1}
(L_2)$ and $(P_{2}^{-1}(K_2), P_{2}^{-1}(L_2))^{\uparrow r}$ is
distributable with respect to $P_{1}^{-1}(L_1)$. Therefore, by
Theorem \ref{theorem10},
\begin{align*}
K^{\uparrow r} \subseteq & K_1^{\uparrow r} \| K_2^{\uparrow r}.
\end{align*}
\hfill$\square$
We observe that unlike mutual normality for computation of supremal normal
sublanguages, mutual $K$-observability depends on the
specification $K$. We might want to replace $K$  by a larger
language $C$ such that $M$ becomes $C$-observable with respect to
$L$. If we insists on having structural conditions that depend on
the the plant only, we need to consider a stronger version of
relative ($C$-)observability, namely with $C=L$ instead of $C=K$.
We recall that $L-$observability of $K$ is still weaker then
normality (because in case of normality the requirement for
unobservable $a$-steps is much stronger). We denote by
$(K,L)^{\uparrow R}$ the supremal sublanguage of $K$ that is
$L$-observable with respect to $L$. Since $L$-observability is
weaker than normality, $(K,L)^{\uparrow R}\supseteq
(K,L)^{\uparrow n}$.

Exchangeability of $(.)^{\uparrow R}$ can be established using the same arguments as Proposition \ref{proposition28}.
Distributivity of $(.)^{\uparrow R}$ is proved below.
\begin{proposition} (Distributivity of $(.)^{\uparrow R}$)
\label{proposition30R}
Let $K,L,M \subseteq A^*$ and $K \subseteq L$. If $M$ is
$L$-observable with respect to $L$, then  $(K,L)^{\uparrow R}$ is
distributable with respect to $M$, that is,
$$
(K \cap M)^{\uparrow R} = K^{\uparrow R} \cap M.
$$
Or,
$$
(K \cap M,L \cap M)^{\uparrow R} = (K,L)^{\uparrow R} \cap M.
$$
\end{proposition}
{\bf Proof}
Since $L$ is closed, $w' \in L \wedge w'a \in L \Leftrightarrow
w'a \in L$. Hence, $M$ is $L$-observable with respect to $L$ if
and only if
 \Lfteqn
 \label{Equation33}
(\forall w \in A^*)(\forall w' \in O^{-1}O(w))(\forall a \in A) \\
wa \in M \wedge w'a \in L \Rightarrow w'a \in M.
 \Ndeqn
For the same reason, $(K \cap M)^{\uparrow R}$ is $(L \cap
M)$-observable with respect to $(L \cap M)$ if and only if
 \Lfteqn
 \label{Equation44}
(\forall w \in A^*)(\forall w' \in O^{-1}O(w))(\forall a \in A) \\
wa \in (K \cap M)^{\uparrow R} \wedge w'a \in (L \cap M) \\
\Rightarrow w'a \in (K \cap M)^{\uparrow R}.
 \Ndeqn
$K^{\uparrow R}$ is $L$-observable with respect to $L$ if and only
if
 \Lfteqn
 \label{Equation55}
(\forall w \in A^*)(\forall w' \in O^{-1}O(w))(\forall a \in A) \\
wa \in K^{\uparrow R} \wedge w'a \in L \Rightarrow w'a \in
K^{\uparrow R}.\\
 \Ndeqn
We now prove the result. \newline
\noindent ($\subseteq$:) Clearly,
$$
(K \cap M)^{\uparrow R} \subseteq K \cap M \subseteq M.
$$
To prove $(K \cap M)^{\uparrow R} \subseteq K^{\uparrow R}$, we
need to show the following. (1) $(K \cap M)^{\uparrow R} \subseteq
K$, which is clearly true. (2) $(K \cap M)^{\uparrow R}$ is
$L$-observable with respect to $L$, that is,
\begin{align*}
&(\forall w \in A^*)(\forall w' \in O^{-1}O(w))(\forall a \in A) \\
& wa \in (K \cap M)^{\uparrow R} \wedge w'a \in L \\
&\Rightarrow w'a \in (K \cap M)^{\uparrow R}.
\end{align*}
This holds because
\begin{align*}
& wa \in (K \cap M)^{\uparrow R} \wedge w'a \in L \\
\Rightarrow & wa \in (K \cap M)^{\uparrow R} \wedge w'a \in L \wedge wa \in M \\
\Rightarrow & wa \in (K \cap M)^{\uparrow R} \wedge w'a \in L \wedge w'a \in M \\
& (\mbox{by Equation (\ref{Equation33})}) \\
\Rightarrow & w'a \in (K \cap M)^{\uparrow R}  (\mbox{by Equation (\ref{Equation44})}).
\end{align*}
\noindent ($\supseteq$:) To prove $K^{\uparrow R} \cap M \subseteq
(K \cap M)^{\uparrow R}$, we need to show that (1)
$K^{\uparrow R} \cap M \subseteq K \cap M$ (clearly true) and
(2) $K^{\uparrow R} \cap M$ is $(L \cap M)$-observable with
respect to $(L \cap M)$, i.e.,
\begin{align*}
&(\forall w \in A^*)(\forall w' \in O^{-1}O(w))(\forall a \in A) \\
&wa \in (K^{\uparrow R} \cap M) \wedge w'a \in (L \cap M) \\
& \Rightarrow w'a \in (K^{\uparrow R} \cap M).
\end{align*}
This is true because
\begin{align*}
& wa \in (K^{\uparrow R} \cap M) \wedge w'a \in (L \cap M) \\
\Rightarrow & wa \in K^{\uparrow R} \wedge w'a
\in L \wedge w'a \in M \\
\Rightarrow & wa' \in K^{\uparrow R} \wedge w'a \in M  (\mbox{ by Equation (\ref{Equation55})}) \\
\Rightarrow & wa' \in (K^{\uparrow R} \cap M) .
\end{align*}
\hfill$\square$
Theorem \ref{theorem26R} follows from Theorem \ref{theorem10}
and Proposition \ref{proposition30R}.
\begin{theorem} (Comparison for $(.)^{\uparrow R}$)
\label{theorem26R}
Given $K_i \subseteq L_i \subseteq A_i^*$,
$i=1,2$, and $K=K_1 || K_2$, $L= L_1 || L_2$. If $P_1^{-1}(L_1)$
is $P_2^{-1}(L_2)$-observable with respect to  $P_2^{-1}(L_2)$ and
$P_2^{-1}(L_2)$ is $P_1^{-1}(L_1)$-observable with respect to
$P_1^{-1}(L_1)$, then
\begin{align*}
K^{\uparrow R} \subseteq & K_1^{\uparrow R} \| K_2^{\uparrow R}.
\end{align*}
That is,
\begin{align*}
(K, L)^{\uparrow R} \subseteq (K_1, L_1)^{\uparrow R} \| (K_2,
L_2)^{\uparrow R} .
\end{align*}
\end{theorem}
Finally we show an example of local plants that satisfy the condition of 
Theorem \ref{theorem26R}, but are not mutually normal.
\begin{figure}
  \label{exmutualRO}
    \centering
    \includegraphics[scale=.57]{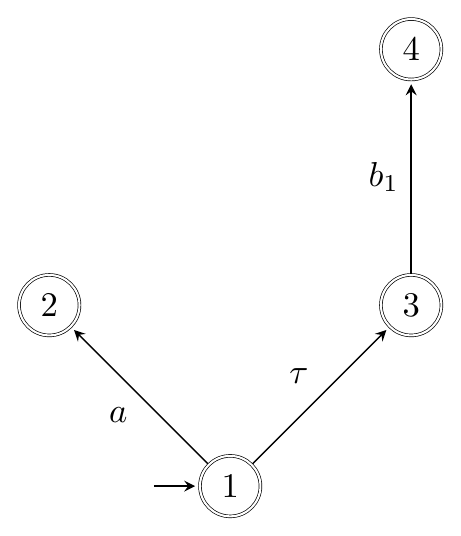}
    \includegraphics[scale=.57]{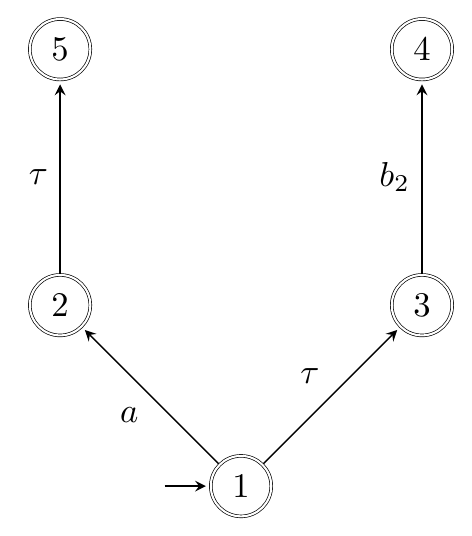}
    \caption{Generators for languages $L_1$ and $L_2$}
    \label{fig4}
  \end{figure}
\begin{example}
\label{exampleRO}
Let $A_1=\{a,b_1,\tau\}$, $A_2=\{a,b_2,\tau\}$,  $A_o=\{a,b_1,b_2\}$, and $A_{uo}=A\setminus A_o$.
Consider local plant languages $L_1$ and $L_2$ on figure 2.
  
Then $L_1$ is not normal with respect to $P_1P_2^{-1}(L_2)$
because $a\tau\in O_1^{-1}O_1(L_1)\cap P_1P_2^{-1}(L_2)$, but $a\tau\not\in L_1 $. 
However, $P_1^{-1}(L_1)$
is $P_2^{-1}(L_2)$-observable with respect to  $P_2^{-1}(L_2)$ and
$P_2^{-1}(L_2)$ is $P_1^{-1}(L_1)$-observable with respect to
$P_1^{-1}(L_1)$. This shows that not only supervisors based on $(.)^{\uparrow R}$ are more permissive 
than those based on $(.)^{\uparrow n}$, but sufficient condition for locally computing the former   is strictly weaker than mutual normality.
\end{example} 
\section{Concluding remarks}
\label{conclusion}

We have studied distributed computation of supremal sublanguages in
modular (and more generally coordination) control of large DES. We
have presented a new approach based on three algebraic conditions
that are key for modular control to be as permissive as monolithic
control. Sufficient conditions for maximal
permissiveness of modular/coordination control are formulated
in terms of global mutual normality and relative
observability. 

In a future research we plan to extend these results to distributed
computation of supremal languages with respect to other properties
closed under unions such as opacity. 

%


\end{document}